**TITLE :**

Behavior of bulk high-temperature superconductors of finite thickness subjected to crossed magnetic fields : Experiment and model

**AUTHOR NAMES :**

Ph. Vanderbemden (1)
Z. Hong, T. A. Coombs (2)
S. Denis (3)
M. Ausloos (4)
J. Schwartz, I. B. Rutel (5)
N. Hari Babu, D. A. Cardwell and A. M. Campbell (6)

**AUTHOR AFFILIATIONS :**

(1) SUPRATECS and Department of Electrical Engineering and Computer Science B28, Sart-Tilman, B-4000 Liège, Belgium

(2) Centre for Advanced Photonics and Electronics, Engineering Department, University of Cambridge, 9 JJ Thomson Avenue, Cambridge CB3 0FA, United Kingdom

(3) SUPRATECS, Communication, Information Systems & Sensors (CISS) Department, Royal Military Academy of Belgium, 30 avenue de la Renaissance, B-1000 Bruxelles, Belgium

(4) SUPRATECS and Physics Institute B5, University of Liège, Sart-Tilman, B-4000 Liège, Belgium

(5) National High Magnetic Field Laboratory, Florida State University, 1800 E. Paul Dirac Drive, Tallahassee, FL 32310, U.S.A.

(6) IRC in Superconductivity, University of Cambridge, Madingley Road, Cambridge CB3 0HE, United Kingdom

# ABSTRACT:

Crossed magnetic field effects on bulk high-temperature superconductors have been studied both experimentally and numerically. The sample geometry investigated involves finite-size effects along both (crossed) magnetic field directions. The experiments were carried out on bulk melt-processed Y-Ba-Cu-O (YBCO) single domains that had been pre-magnetized with the applied field parallel to their shortest direction (i.e. the *c*-axis) and then subjected to several cycles of the application of a transverse magnetic field parallel to the sample *ab* plane. The magnetic properties were measured using orthogonal pick-up coils, a Hall probe placed against the sample surface and Magneto-Optical Imaging (MOI). We show that all principal features of the experimental data can be reproduced qualitatively using a two-dimensional finite-element numerical model based on an *E-J* power law and in which the current density flows perpendicularly to the plane within which the two components of magnetic field are varied. The results of this study suggest that the suppression of the magnetic moment under the action of a transverse field can be predicted successfully by ignoring the existence of flux-free configurations or flux-cutting effects. These investigations show that the observed decay in magnetization results from the intricate modification of current *distribution* within the sample cross-section. The current *amplitude* is altered significantly only if a field-dependent critical current density $J_c(B)$ is assumed. Our model is shown to be quite appropriate to describe the cross-flow effects in bulk superconductors. It is also shown that this model does not predict any saturation of the magnetic induction, even after a large number (~ 100) of transverse field cycles. These features are shown to be consistent with the experimental data.

# 1. INTRODUCTION

The magnetic properties of hard, type-II superconductors containing an array of parallel vortices may be understood within the framework of the Bean critical state model.[1] The situation becomes more intricate for a non-parallel flux line lattice, however, that may arise, for example, when a superconductor is subjected to a rotating magnetic field or to a magnetic field applied perpendicular to its magnetization. The latter is referred to as a "*crossed magnetic fields*" configuration. Despite more than three decades of investigation,[2-4] the magnetic behavior of bulk type II superconductors under crossed magnetic fields remains an intriguing, but relatively ill-understood, topic. Moreover the interest in crossed fields has been revived with the increasing potential of single domain bulk melt-processed RE-Ba-Cu-O ((RE)BCO, where RE denotes a Rare Earth ion) to trap significant magnetic fields[5], which gives rise to the possibility of high-field permanent magnet-like devices.[6,7] Applying a magnetic field in a direction transverse to that of the pre-magnetization in a bulk superconductor may cause significant decay of the trapped field,[8-18] which could result in the failure of such devices in practical applications. The behavior of bulk HTS in the crossed field configuration is therefore of practical relevance as well as of fundamental academic interest.

Early experimental investigations on crossed field configurations involved measuring the magnetic moment of type-II low-$T_c$ superconducting wires subjected to both axial and transverse magnetic fields.[2-3] Funaki and Yamafuji investigated the behavior of low-$T_c$ materials in the presence of mutually perpendicular DC and AC magnetic fields.[19] In some geometries, the results suggested that the applied AC field suppressed the DC pre-magnetization to varying degrees. Similar effects were observed in high-$T_c$ superconductors, initially for cylindrical YBCO carrying a transport current and subjected to a magnetic field parallel to the cylinder axis[20] and subsequently in high-$T_c$ superconductors of various shapes and microstructures.[8-18,21] It is now well-established experimentally that the magnetization $M_z$ of a type-II superconductor sample placed in an external DC magnetic field $H_z$ is decreased strongly by applying a magnetic field $H_y$ perpendicular to both $M_z$ and $H_z$. This behavior is known as *collapse of magnetic moment*.[12] It has also been observed that subsequent sweeps of the transverse field $H_y$ give rise to further reductions of $M_z$ to a value corresponding to the limiting case for which $H_y$ is an AC magnetic field applied orthogonally to the DC field.

Several theoretical approaches attempt to describe the experimental observations, including the pioneering works of Bean[22] or the double critical-state model of Clem.[23,24] This model includes the effects of both flux-line pinning and flux-line cutting, i.e. the cross joining of non-parallel vortices at their intersection. In order to further improve the agreement with experimental data, Fisher *et al.* developed a two-velocity hydrodynamic model involving two vortex systems moving at different speeds.[12,14] This model, however, is limited to small variations of the tilt angle of the total magnetic field,[25] and, as a result, an elliptic flux-line-cutting critical-state model was developed subsequently.[26,27] In these approaches, the length of the sample is assumed infinite in the direction of both crossed fields, which only arises either for (i) an infinite slab with crossed magnetic fields parallel to the slab surface or (ii) an infinite cylinder subjected to both azimuthal and axial fields.

Another technique employed to investigate cross-flow effects in superconductors is based on a variational approach. Bhagwat *et al.*,[28] for example, proposed a method based on minimizing the total flux change in the sample. This variational approach was generalized by Badía *et al.* who modeled the electrodynamic response of the superconductor by minimizing an appropriate functional related to magnetic field changes within the material.[29-31] In this work, the boundary conditions are determined by the external magnetic field source and the two components of the local current density vector are such that its extremity is constrained to "stay" on some locus defined by the geometry-dependent critical state model employed. Such geometries include (i) a disk of constant radius (isotropic Bean model), (ii) a rectangle (double critical state model), (iii) an ellipse (elliptic model), or (iv) a disk whose radius is a function of the magnetic field angle with respect to a reference axis ("pseudo-isotropic model").[31] Recently, such variational approaches have been applied successfully to samples of finite size.[32,33] To our knowledge however, the studies carried out up to now in the "crossed-field" configuration do not take demagnetization effects into account: the sample geometry is assumed to be infinitely long in the direction of the (crossed) applied fields.

The case of thin strips in the presence of a cross-flux interaction was first studied using a model to calculate the attractive interaction core between two orthogonal vortex lattices in a layered superconductor.[34] Recently, Brandt and Mikitik have analysed successfully the behavior of a thin superconductor in the crossed field configuration.[35-37] The sample was considered initially to consist of a thin superconducting strip (width $w$, thickness $d \ll w$) subjected to both a transverse DC magnetic field ($H_{DC} \perp$ strip) and an AC magnetic field $H_{AC}$ perpendicular to the DC field (i.e. parallel to the plane of the strip). Semi-analytical expressions were obtained in the two cases where $H_{AC}$ is either parallel ("*longitudinal vortex shaking*") or perpendicular ("*transverse vortex shaking*") to the currents circulating in the strip. The theories were extended subsequently to describe rectangular platelets (dimensions $w$, $L$ << thickness $d$).[37] The geometry assumed is such that currents are always perpendicular to the field. It predicts that for an AC amplitude less than the penetration field in the direction normal to the trapped moment, the moment will stabilize, while for amplitudes greater than this it will disappear completely.[35]

The theoretical techniques described above have the common characteristic of assuming a *vertical E-J* characteristic: infinite dissipation is supposed to occur when the critical current components exceed some given value. An alternative approach involves modeling the superconductor with a highly *non-linear E-J* constitutive law $E \propto J^n$, with $n$ being a large number.[38-40] Such a relation can be incorporated within commercial Finite Element Method (FEM) software used for simulating physical processes based on Partial Differential Equations (PDE). Within this context, a general method of analyzing numerically the electromagnetic behavior of high-temperature superconductors subjected to time-varying magnetic fields was proposed recently by Hong *et al.*[41-44] A commercial FEM modeling package was used to model the two-dimensional magnetization of bulk superconductors of various shapes of finite size in the direction of the applied magnetic field. In numerical approaches of this type, the electric field $\vec{E}$ is always assumed to be parallel to the current density $\vec{J}$.

The objective of the present work is to investigate the validity of the $\vec{E} \parallel \vec{J}$ approach for modeling the results of the crossed field experiments on bulk high-temperature superconductors. Despite its extreme simplicity, this approach will be shown to reproduce successfully many features of the collapse of magnetic moment under transverse fields. In addition to being relevant to several applications of bulk HTS magnets, such a geometry will also enable the spatial distribution of the *c*-axis magnetic flux on the top surface of superconductor to be predicted before and after transverse fields are applied and to compare theoretical predictions to experimental data.

This paper is organized as follows. The details of the experiment are given in section 2. In section 3, the method used for modeling the electromagnetic behavior of the sample is described. Section 4 compares the modeled data to the experimental results. First, the behavior of the global remanent magnetization of a sample subjected to a series of transverse field cycles is studied in detail. Next, the model is used to compute the local magnetization on the top surface of the sample and to compare these with Magneto-Optical Imaging (MOI) data from experiments on bulk single domains. In section 5, the validity and limitations of the proposed model are discussed and compared to other approaches. Finally, conclusions are drawn from the analyses given in the preceding sections.

## 2. EXPERIMENT

### A. Sample preparation and characterization

Bulk melt-textured $YBa_2Cu_3O_{7-x}$ (YBCO) and $DyBa_2Cu_3O_{7-x}$ (DyBCO) single domains were fabricated by a top seeded melt growth (TSMG) technique, as described in Refs. 45-48. The as-grown samples were cylindrical with a typical diameter of 20 mm and thickness of 10 mm (i.e. parallel to the $c$-axis). Here, the focus is on measurements carried out on two samples: (i) a whole cylindrical single domain (sample HB1) and (ii) a small parallelepipedic sample (HB2) cut from a whole single domain using a wire saw and intended for Magneto-Optic Imaging measurements. Sample dimensions are summarized in Table I. Both samples have an aspect ratio ~ 1:3 and their faces coincide with the crystallographic planes of the (RE)BCO material.

Due to the large size of sample HB1, classical magnetic moment measurement techniques (such as SQUID, VSM) [49,50] could not be used to characterize the sample. Instead, two orthogonal pick-up coils wound closely around the sample were used simultaneously to record components of the magnetic induction along the $c$ and $ab$ directions using an applied magnetic field sweep rate of $\mu_0 dH/dt$ ~ 5 mT/s. The pick-up coil voltages were amplified using a SR560 low-noise pre-amplifier, measured using a Keithley 2001 voltmeter and integrated numerically to obtain the average magnetic induction in each direction. Prior to measurements in the crossed field configuration, the magnetization loops $M(H)$ of each sample were measured with the field applied in one given direction, i.e. either $H \parallel c$ or $H \parallel ab$. The magnetic characterization of sample HB2 was carried out in a Quantum Design Physical Property Measurement System. The full-penetration fields $\mu_0 H_p$ of both samples determined for $H \parallel ab$ and $H \parallel c$ are summarized in Table I. Magnetization loops recorded at low fields for both directions revealed a monotonically decreasing $J_c(B)$ for both samples.[50,51] The anisotropy ratio $J_c(ab) / J_c(c)$ was found to be close to 1.5- 2 in the temperature and magnetic field range investigated.

**B. Measurements in the crossed field configuration**

In all crossed field experiments, the melt-textured samples were first pre-magnetized parallel to the *c*-axis by field cooling (FC) to the target temperature (either $T = 77$ K or $T = 80$ K) under an applied field of either 0.5 T (sample HB1) or 0.12 T (sample HB2). According to the $H_p$ values listed in Table 1, such fields were sufficient to generate the maximum remanent magnetization in the samples. The *c*-axis field was then removed and a constant time interval (two minutes) allowed for magnetic relaxation. A series of transverse magnetic field cycles were then applied parallel to the *ab* plane of the samples. A schematic illustration of the measurement configuration is shown in Figure 1(a). The directions of the remanent magnetization ($\parallel c$) and of the transverse applied fields ($\parallel ab$) will be denoted hereafter by *z* and *y*.

A pick-up coil wound closely around the sample was used to measure the average magnetic properties along the *z*-axis for sample HB1 as described above. In addition, an AREPOC Hall probe (active surface 2 x 3 mm$^2$) was placed against the sample surface in order to measure the induction parallel to the *z*-axis in the central zone of the pellet. A previous study[17] showed that the magnetization collapse curves measured by the Hall probe or the sensing coil display similar qualitative behavior. Prior to performing any measurement sequence, the angular position of the sample was adjusted carefully above $T_c$ in order to ensure that the transverse fields were strictly orthogonal to the *z*-axis. The sample was then clamped firmly to prevent rotation by the relatively strong magnetic torque generated by the transverse fields during the experiment.

The spatial distribution of the trapped magnetic induction parallel to the *z*-axis for the smaller sample (HB2) was recorded using a Magneto-Optic Imaging (MOI) system.[52-57] In this measurement, a Bi-substituted ferrimagnetic iron garnet (BIG) indicator thin film with in-plane magnetization and exhibiting a large Faraday effect was placed in contact with the large face of the sample (i.e. perpendicular to the *z*-axis). The sample was placed in a Janis, liquid

He optical cryostat attached to the xyz stage of an optical polarizing microscope (Olympus BXFM) connected to a CCD camera. Our MOI system is sensitive to the modulus of the normal component of the magnetic induction $|B_z|$. Our set-up can reveal the spatial distribution of $|B_z|$ with typical magnetic and spatial resolutions of ~ 10 mT and ~ 10 μm respectively. A concentric air coil was used around the cryostat to generate a field $\mu_0 H \| z$ up to 120 mT. A pair of Helmoltz coils was used to generate the transverse fields $\mu_0 H \| y$, ranging between -34 and +34 mT.

In addition to being sensitive to the distribution of the *normal* magnetic induction $B_z$, the MO film is also weakly sensitive to the magnetic induction component *parallel* to the film $B_y$.[58,59] It is important to ensure that the contribution of the transverse field to the MO signal is negligible if the MOI technique is to be used effectively for crossed field experiments. A detailed study[58] shows that the gray level values, $G$, plotted in a MO image can be expressed as a function of $B_z$ and $B_y$ as

$$G \propto \frac{B_z^2}{\left(B_y + B_A\right)^2 + B_z^2},$$

where $B_A$ is a parameter of the MO indicator film ($B_A = E_A / M_s$, where $E_A$ is the anisotropy energy and $M_s$ the spontaneous magnetization). The calibration of our MO film with superimposed $B_y$ and $B_z$ inductions of known amplitude yields $B_A \sim 144$ mT. In our experiments, the MO images were recorded either *before* or *after* applying the transverse field; as a result the *applied* field was zero. However, a longitudinal component of the transverse field is to be expected from the sample magnetization along the *y*-axis. From the results of the model presented below, a maximum estimate of the transverse induction $B_y$ along the sample top surface was made to be 4.2 mT. The corresponding error on the determination of $B_z$ is less than 3 %.

## 3. MODEL

The numerical method used for modeling the electromagnetic behavior of a bulk superconductor was based on solving the set of Maxwell equations in two dimensions using the Finite-Element Method (FEM) software *Comsol Multiphysics 3.2*. The numerical scheme has been published previously[41] and can be summarized as follows. The space is divided into two sub-domains: the superconducting region and the air. A set of Partial Differential Equations (PDEs), sharing the same dependent variables, is defined in each sub-domain. By assuming that the constitutive law $\vec{B} = \mu_0 \vec{H}$ applies both in air and in the superconducting region, the relevant Maxwell equations are

$$\vec{\nabla} \times \vec{E} = -\mu_0 \frac{\partial \vec{H}}{\partial t} \qquad (1)$$

$$\nabla \times \vec{H} = \vec{J} \qquad (2)$$

The *E-J* behavior of the superconducting material is modeled assuming (i) $\vec{E} \parallel \vec{J}$ and (ii) a power law relationship, i.e.

$$\vec{E} = E_c (\vec{J}/J_c)^n \qquad (3)$$

where $E_c$ denotes the threshold electric field used to define the critical current density $J_c$. We used $E_c = 1\ \mu V/cm$ and a constant $n = 21$, in agreement with typical values for bulk melt-processed YBCO.[60,61] The critical current density can be either constant or dependent on applied magnetic field. If $J_c$ is field dependent, we use Kim's law[62]

$$J_c(|B|) = \frac{J_{c1}}{1 + \frac{|B|}{B_1}}, \qquad (4)$$

where $|B|$ denotes the modulus of the local magnetic induction; $J_{c1}$ and $B_1$ are constant parameters. In contrast with Ref. 63 where FEM Modelling was carried out using an anisotropic version of the Kim model, the critical current density is supposed to be independent of the direction of magnetic field in the present work.

In the two-dimensional model used here, the space is assumed to be infinite in the *x* direction (perpendicular to the plane of the paper), as shown in Figure 1(b). The sample is assumed to consist of an infinite cylinder of rectangular cross-section $y_0 \times z_0$, i.e. filling the space $|x| < \infty$, $|y| < y_0/2$, $|z| < z_0/2$. In this geometry, the magnetic flux lies in the *yz* plane; the current density lines $\vec{J}$ flow in the *x* direction only and close at infinity. By substituting Eqs.

(3) and (2) into (1), two PDEs for two variables $H_y$ and $H_z$ are obtained, representing the components of the magnetic field in the *y* and *z* directions, respectively. A Dirichlet boundary condition is used at infinity whereas the continuity equation $\vec{n} \times (\vec{H}_1 - \vec{H}_2) = \vec{0}$ is used at the boundary between air and superconductor. The PDEs are solved using the FEM software, subject to the boundary conditions.

Although the two-dimensional geometry used in the model does not represent the true geometry of the superconducting samples used in experiments (see Figures 1(a) and 1(b)), the key point is that the sample has *finite* dimensions along both the remanent magnetization (*z*) and the applied transverse field (*y*) directions. In order to simulate the true cross-section of the samples measured here (cf. Table 1), the dimensions used in the model were chosen always such that $y_0 = 3\ z_0$. The two-dimensional modeling could not take the actual $J_c$ anisotropy into account, however, since the current is assumed to flow in only one direction. However the parameter $B_1$ appearing in Kim's law (Eq. (4)) was chosen such that the *M(H)* curves modeled in both directions could reproduce the experimental data in a qualitative manner. Taking $B_1 \sim 0.25$, $\mu_0 H_p \parallel c$ was found to lead to very satisfactory results.

The superconductor must be pre-magnetized parallel to the *z*-axis prior to the application of crossed field. This was achieved by zero field cooling (ZFC) and then subjecting the sample to an increasing magnetic field, ramped linearly for 0.05 s to some value exceeding twice the full-penetration field.[64] $H_z$ is then decreased towards zero at the same sweep rate. A constant time interval (10 seconds) was then employed in order to allow the trapped magnetic moment to relax due to flux creep effects.[65-67] A series of magnetic field cycles parallel to the *y* direction was then applied to each sample. Finally, the program is used to compute the two magnetic field components $H_y$ and $H_z$, from which are determined (i) the current distribution within the sample cross-section, (ii) the *local* magnetic field $H_z$ at the top surface of the sample and (iii) the *global* sample magnetic moment in the *z* and *y* direction.[41]

Table II summarizes the modeling parameters that are either fixed or variable in the framework of the present study.

## 4. RESULTS

The results of three different crossed field experiments are presented in this section. These data are then compared with the predictions of the model. Initially, the "collapse" of the trapped magnetization caused by *one* cycle of the transverse field is investigated, and the influence of the transverse field amplitude examined. Secondly, the effect of a *large number* of transverse field sweeps is considered. Thirdly, the remanent magnetic flux *distribution* above the top surface of the sample before and after applying the transverse field sweeps is reported. The final part of this section compares the results of the model with those obtained for "paramagnetic" and "diamagnetic" initial states of the magnetization $M_z$, i.e. when the transverse field $H_y$ is cycled in the presence of a static field $H_z$, which is either parallel or anti-parallel to $M_z$.

**A. Magnetization collapse for one sweep of the transverse field**

Figure 2 shows the influence of transverse magnetic field sweeps $H_y$ on the trapped magnetic induction $B_z$ (|| *c*-axis) measured by a Hall probe placed at the centre of sample HB1. The magnetic induction is normalized with respect to its initial value, $B_0$, whereas the transverse field is normalized with respect to the full-penetration field $H_p$ in the *y* direction (i.e. for $H$ || *ab*). The experiment was carried out for several transverse field amplitudes $H_{max}$ that are smaller, equal or larger than $H_p$. The experimental data display the typical features observed in similar experiments.[8,9,12,14] Firstly, as the transverse field amplitude increases, the remaining induction after one complete transverse field cycle decreases; for $H_{max} \sim H_p$, the remaining $B_z$ after one cycle is found to be ~ 0.65 $B_0$. Secondly, the consistent decreases in $B_z$ observed when $H_y$ is swept from 0 to $+H_{max}$ and from 0 to $-H_{max}$ (plain line in the inset of Fig. 2) are generally much larger than the decrease in magnetic induction caused by sweeps from $+H_{max}$ to 0 and from $-H_{max}$ to 0 (dashed line in the inset of Fig. 2). Thirdly, the results displayed in Figure 2 also show that cycling the transverse field always causes the induction to decrease, even at very small amplitudes down to 0.15 $H_p$.

Figure 3 shows the results of the model, corresponding to the experimental conditions described above. The transverse field $H_y$ is swept for 0.4 s and a field-dependent $J_c(B)$ is assumed. From a quantitative point of view, the induction decays resulting from one complete oscillation of the transverse field are found to be much more pronounced than those measured experimentally: after one cycle of amplitude $\sim H_p$, the remaining $B_z$ is only $\sim 0.10\ B_0$, compared to ~ 0.65 $B_0$ for the experimental data. From a qualitative point of view, however, the modeled $B_z(H_y)$ curves are remarkably similar to the experimental data. In particular, the $B_z(H_y)$ segments that correspond to sweeps from $+H_{max}$ to 0 and from $–H_{max}$ to 0 (dashed line in the inset of Fig. 3) are nearly horizontal, showing clearly that the remanent magnetization is affected only weakly by the transverse field in this regime. We also notice that small amplitudes of the transverse field (down to 0.15 $H_p$) lead to a *monotonic* decrease of the magnetic induction, as observed experimentally.

Figure 4 shows the current distribution modeled within the sample cross-section at several selected times during the first transverse field cycle for transverse field amplitude of 0.5 $H_p$. The parameters of the model are identical to those used in Fig. 3, i.e. we consider a field dependent $J_c(B) = J_{c1} (1 + 4|B|/B_p)^{-1}$, where $B_p$ is the full-penetration induction. Image 0 shows the current distribution that gives rise to a positive trapped magnetization (i.e. $M_z > 0$). The magnitude of the current density is not uniform within each half of the cross-section. This arises from the $J_c(B)$ dependence: during the magnetization process, the local flux density $|B|$ varies between 0 and 2 $B_p$, which corresponds to critical current densities ranging between $J_{c1}$ and $J_{c1}/9$. The application of a positive transverse field $H_y > 0$ leads to a reversal of the current density in the top-left and bottom-right regions of the sample, as shown in Image 1. In other words, a thin horizontal layer located at the top (resp. the bottom) of the sample carries negative current (resp. positive). These current directions correspond to those required for shielding the increasing applied field $dH_y/dt > 0$. The consequence is that both upper and lower layers of the sample no longer contribute efficiently to the $z$-axis magnetization $M_z$. Increasing the field (Image 2) increases the distortion of the current distribution. On lowering the field (Images 3 and 4), the shielding currents in the top and bottom layers change their sign, as required for shielding the decreasing applied field $dH_y/dt < 0$. However the current distribution in the central horizontal layer of the sample (i.e. responsible mainly for the $z$-axis magnetization) remains unaffected. The corresponding decay of $M_z$ is therefore very small, as observed in Figures 2 and 3. As the applied field is decreased further (Images 5 and 6), the top and bottom layers carrying positive and negative currents, respectively, extend further towards the middle plane ($z = 0$) to the detriment of the central layer. This results in a significant decrease of $M_z$. In contrast, changes of the current distribution arising during the final part of the transverse field cycle (Images 7 and 8) relate primarily to the upper and lower layers of the sample. This results in an almost unaffected $M_z$, as observed above. In addition to affecting the current *distribution* within the sample, the transverse field cycles also affect the *magnitude* of the current density, as can be shown in Figure 4. One reason is likely that the transverse field penetrating from the top and bottom sides of the sample affects the flux line density $|B|$ and, accordingly, the value of $J_c$.

In order to discriminate between the effects due to the $J_c(B)$ dependence and those arising only from the transverse field, the same model was used but with a constant $J_c$. The results are shown in Fig. 5(a) for the first half of the cycle $0 \rightarrow H_{max} \rightarrow 0$, with $H_{max} = 0.5\, H_p$. Figures 5(b) and 5(c) compare the current distributions for both constant and field-dependent $J_c$ when $H_{max} = 1.5\, H_p$. The resulting average magnetization $M_z(H_y)$ curves in each case are summarized in Fig. 5(d). These indicate minimal *qualitative* difference between the "constant $J_c$" or "field-dependent $J_c$" hypotheses, although the magnetization collapse is always stronger for a field-dependent $J_c$. The current distributions shown in Figs. 4 and 5(a) display similar "front lines", separating positive and negative values of $J_c$. The magnitude of current, however, is very inhomogeneous when a $J_c(B)$ dependence is assumed, unlike the case for "constant $J_c$". At high transverse field amplitudes, the differences between the two current distributions are much more marked (Figs. 5(b) and 5(c)): for the "$J_c(B)$ case", the current amplitude is reduced significantly when the transverse field reaches significant values (Images (1)-(3) in Fig. 5(c)). In this regime, the flux line density $|B| = (B_y^2 + B_z^2)^{1/2}$ is large, resulting in a significant decrease of $J_c$. The sample reacts to reduce the induced electric field $E = E_c\, (J/J_c)^n$ by reducing the current density amplitude $J$ without changing the current distribution, as observed by comparing Images (1) and (2) in Fig. 5(c).

In summary, the results of the model show that the application of a transverse field $H_y$ affects both (i) the current *distribution* in both top and bottom parts of the sample and (ii) the current *amplitude* at every point of the sample when a $J_c(B)$ dependence is assumed. Both effects lead to a decrease of the *z*-axis magnetization $M_z$. It should be noted that additional modeling was carried out for a sweep rate amplitude of over two decades, i.e. by sweeping one cycle of the transverse field for 0.04 s, 0.4 s and 4 s. This resulted in minor modifications of the $M_z(H_y)$ plots. Similarly, a modification of the value of $n$ used in the modelling ($n = 15, 21, 25, 31$) has only little effects on the results, as expected from scaling arguments.[39,68] Therefore, even though a finite *n*-value is assumed, the sweep rate is not considered a relevant parameter affecting the results in the *E-J-B* conditions involved in the present study.

**B. Magnetization collapse for a large number of sweeps of the transverse field**

The influence of a large number of transverse field sweeps on the remanent induction at the centre of the sample is examined next. The experimental results are presented in Fig. 6, which shows the normalized magnetic induction $B_z$ measured by the Hall probe at the end of each transverse field cycle of single polarity > 0, i.e. $0 \to H_{max} \to 0 \to H_{max} \to 0$ etc. Successive cycles cause the magnetic induction to decrease by smaller and smaller amounts, as illustrated by the log-log plot of the data. The striking feature of Fig. 6 is that the induction $B_z$ does not appear to saturate, even after a large number of field sweeps. Indeed, the $B_z$ vs. cycle number $N$ curves can be fitted using a power law $B_z \sim N^{-\alpha}$, with the exponent $\alpha$ being an increasing function of the transverse field amplitude $H_{max}$.

Similar experimental conditions were used in the modeling, and the results are shown in Figure 7. Good agreement between the experimental data and a power law fit is observed over the whole range of $N$ ($1 < N < 100$). As is observed experimentally, the remanent induction continues to decrease after 100 sweeps, and no saturation occurs.

## C. Magnetic flux distribution measurements

The influence of a transverse magnetic field on the trapped magnetic flux distribution was investigated by Magneto-Optical (MO) imaging. Initially, the trapped-flux profile was recorded after having magnetized the sample permanently along the $z$ direction (i.e. || $c$-axis). A typical MO image of the magnetic flux distribution $B_z(x,y)$ above sample HB2 at $T = 80$ K is shown in Fig. 8(a), in which the brighter-in-contrast regions correspond to higher values of $|B_z|$. The image shows a regular trapped-field pattern of nearly pyramidal shape, as expected from similar experiments on melt-textured materials.[69-71] A transverse field $H_y = 0.48\, H_p$ (|| $ab$) was then applied along the positive $y$ direction (i.e. $H_y > 0$), removed, and the flux distribution re-measured. This procedure was followed by a similar second cycle but in the opposite direction (i.e. $H_y < 0$).

Figure 8(b) shows the distribution of the modulus of the trapped magnetic induction, $|B_z|$, along the $y$ axis, i.e. parallel to the transverse field. It should be noted that all flux profiles display a slightly asymmetric structure. This phenomenon may arise from small inhomogeneities in $J_c$ within the single domain and will not be considered further. The initial profile (black symbols) displays the typical features of the trapped flux profiles measured on similar materials.[70-72] A particular feature is the slight change of curvature on both sides of the peak, which results in a sharper profile in the central zone (–0.2 < $y$ < 0.2 mm). This can be understood by taking into account the very inhomogeneous current distribution associated with the $J_c(B)$ dependence (c.f. Image (0) in Fig. 4). The apparent $|B_z|$ minima that occurs for $y$ = –0.8 and +0.8 mm arise from a change of sign of $B_z$. Note that the $y$ axis used in Figure 8(b) ranges from -0.89 to 0.89 mm, which corresponds to the side length of the sample (1.78 mm). The existence of a negative magnetic induction in the outer perimeter of the sample (i.e. within a layer of ~ 0.1 mm thickness along the edges) is a geometrical effect resulting from the closure of the pinned magnetic flux lines in a sample of finite thickness (0.56 mm in the present case).

The flux profiles recorded after applying the transverse fields are also shown in Fig. 8(b). Careful examination of Fig. 8(b) shows that the transverse fields lower the trapped magnetic induction in a slightly asymmetrical manner; the profiles are shifted to the left and to the right after the first ($H_y > 0$) and the second ($H_y < 0$) field sweeps, respectively. A more visible difference appears around the sample edge in the regions indicated by the arrows in Fig. 8(b); the $|B_z|$ signal after the first sweep increases on the left side of the sample ($y$ ~ – 0.85 mm) and nearly vanishes on the right side ($y$ ~ – 0.85 mm). As a result, the plot appears to "tilt" slightly with respect to the initial data. The second sweep yields the opposite phenomenon. These differences are very small but clearly perceptible from the experimental data.

Figure 8(c) shows the modeled magnetic induction distribution $B_z(y)$ above the top surface of the sample before and after application of crossed fields of amplitude (0.5 $H_p$), which corresponds to that used experimentally. In order to compare the modelled and experimental data, the *modulus* of the induction, $|B_z|$, is shown in Fig. 8(c). The initial profile (black symbols) is symmetric with respect to the $y$ = 0 axis and displays two features that are similar to the experimental data in that: (i) two symmetric inflexion points delimit a slightly sharper profile in the central zone and (ii) $B_z$ changes its sign in an external zone. The profiles recorded after the transverse field sweeps (white symbols) also show noticeable similarities with the experimental data. Firstly, the magnetic flux distributions are seen to be shifted to the left (resp. right) after the positive (resp. negative) field sweep. Secondly, the external parts of the profile are modified significantly, resulting in an asymmetric $B_z(y)$ curve.

### D. Comparison of paramagnetic and diamagnetic initial states

The above results always relate to an initial state for which a pre-magnetization $M_z$ is trapped in the $z$-direction in the absence on an applied $H_z$ field. The model results presented in Fig. 9 compare the results obtained for "paramagnetic" and "diamagnetic" initial states of the magnetization $M_z$, i.e. when the transverse field $H_y$ is cycled *in the presence* of a static field $H_z = H_p$, which is either parallel ($M_z > 0$) or anti-parallel ($M_z < 0$) to the direction of magnetization. As may be seen in Fig. 9, the model predicts that the $M_z(H_y)$ plots in the presence of $H_z > 0$ or $H_z < 0$ are perfectly *symmetric* with respect to each other. The current distribution within the sample cross-section (not shown here) in the presence of a finite $H_z$ component superimposed to the transverse field is quantitatively similar to the current distribution shown in Figs. 4 and 5.

## 5. DISCUSSION

In this section, the applicability and limitations of the model used to simulate the experimental data are discussed. First, the impact of the finite $n$-value is considered. Then, the validity of the two-dimensional geometry for studying cross-flow effects on trapped-field bulk superconducting magnets is discussed. Finally, alternative explanations are proposed for the observed collapse of magnetic moment as a function of applied transverse field by considering the similarities between the results of the model and experiment.

### A. Validity of a *E-J* power law with finite *n*-value

Unlike existing models, which aim to calculate the true critical state in type-II superconductors with cross-flow effects,[12,14,26-31,34-37] the approach adopted here simply considers a $E \propto J^n$ power law (with $n = 21$ in the present case). One of the reasons for the good agreement between the predictions of the model and the experimental results is that the $n$-value in *real* HTS is *finite* and probably very close to 21 for melt-textured YBCO at 77 K, although there is significant discrepancy between $n$ values determined by different experimental techniques. Using a direct (transport) measurement method, Noudem *et al.* reported $n = 7$ in self-field at 77 K.[61] Using indirect (magnetic) measurements at several sweep rates and a suitable model taking into account the current density distribution in the sample,[73] on the other hand, Yamasaki *et al.* observed $n$ values at 77 K ranging between 27.2 ($B$ = 0.5 T) and 31.9 ($B$ = 0.2 T).[60] A power-law model is therefore more suitable than "true critical state" models ($n \to \infty$) in the case of bulk YBCO at liquid nitrogen temperature ($T / T_c$ ~ 0.8).

The consequence of a finite $n$-value is the relaxation of the magnetization due to flux creep.[60,65-67,74-76] It is important, therefore, to ensure that the magnetization decay caused by flux creep is much smaller than the $M_z(H_y)$ collapse caused by transverse fields. In the present model, with $n$ = 21, the decay of the average $M_z$ with time $t$ after the magnetizing field $H_z$ is removed was found to follow the law

$$M(t) = M_0 \left[ 1 + \left( \frac{t}{t_0} \right) \right]^m \qquad (5)$$

($m$ < 0), as is usually observed in trapped field magnets.[77] Flux creep theory[60,65,78] predicts that the $m$ exponent is related to the $n$-value of the $E$-$J$ law by

$$m = \frac{1}{1-n}. \qquad (6)$$

Therefore a theoretical value of $m$ equal to –0.05 is expected in the present case ($n$ = 21). The values of $m$ determined by fitting our modelled $M(t)$ using Eq. (5) are $m$ = –0.05 for a constant $J_c$ and $m$ = –0.034 for a field-dependent $J_c(B)$. The reason for the significant disagreement in the case of a field-dependent $J_c$ may be understood relatively easily, as follows. In order to simplify the algebra, we first replace our $J_c(B)$ dependence $J_c = J_{c1} (1 + 4|B|/B_p)^{-1}$ by a power law

$$J_c = J^* (|B|/B^*)^{-\gamma} \qquad (7)$$

The exponent $\gamma$ is determined by fitting the $J_c(B)$ law in the interval $B_{min} < |B| < B_{max}$, where $B_{min}$ and $B_{max}$ denote the minimum and maximum local values of the local magnetic induction $|B|$ in the sample. In the present case, one has $B_{min} \approx 0.1 B_p$ and $B_{max} \approx B_p$, and the best fit yields $\gamma \approx 0.55$. Incorporating Eq. (7) into $E = E_c (J/J_c)^n$, the *local* electric field $E$ is proportional to $B^{n\gamma} J^n$. Assuming that the average flux density <$B$> is proportional to the average current density <$J$> and that the current density variations are small enough to be neglected to first order, the *average* electric field <$E$> can be expressed as

$$<E> \propto <B>^{n\gamma+n}. \qquad (8)$$

The average electric field is also related to <*B*> through Faraday's law by

$$\langle E \rangle \propto -d\langle B \rangle/dt. \qquad (9)$$

By integrating Eqs (8) and (9), the average flux density <*B*> can be expressed as

$$\langle B \rangle(t) = B_0 \left[1 + \left(\frac{t}{t_0}\right)\right]^m \text{ with } m = \frac{1}{1 - n(1+\gamma)}, \qquad (10)$$

where $t_0$ and $B_0$ are constants. Using $n = 21$ and $\gamma = 0.55$, we obtain $m = -0.032$, which is quite close to the –0.034 value found by fitting the decay of the average magnetization. This simple result is particularly significant and shows how the field-dependence of $J_c$ affects the decay of trapped flux due to the effects of magnetic relaxation.

By neglecting the small changes in the relaxation rates in the "crossed flux" configuration with respect to the "unidirectional field" configuration,[13,79] the contribution of the decay arising from flux creep effects during a crossed field experiment can be estimated. The present model allows a time of 10 s to elapse after the magnetization process. The magnetization decreases by approximately 29 % during this period, and then by a further 0.13 % during the application of one full cycle of the transverse field. Considering a large number (100) of transverse field sweeps (cf. Fig. 7), the corresponding decrease is 4.5 %. Such values are much smaller than the decay in magnetization caused by the transverse fields.

From the above study it is concluded that an *E-J* power law model is quite appropriate to describe the cross-flow effects in bulk superconductor, trapped field magnets.

**B. Validity of the geometry used in modeling**

The two-dimensional approach proposed here assumes that the current density lines $\vec{J}$ flow in *one direction* only and close at infinity. Such a procedure avoids the need to consider two critical current densities $J_{c\perp}$ and $J_{c\parallel}$, as required by flux cutting theory,[23,24] the two-velocity hydrodynamic approach[12,14] or more sophisticated models.[26,27] In the present model, the electric field $\vec{E} \parallel \vec{J}$ is never parallel to the magnetic induction $\vec{B}$. This allows one to describe the magnetic behavior of the material using only *one* parameter $J_c$, i.e. the conventional depinning current density which is directly accessible via experiment.

The validity of the two-dimensional model to describe cross-flow effects in parallipipedic or short cylindrical samples is open to debate. The results seen in Figs. 4 and 5 show clearly that the transverse field penetrates simultaneously from the top and the bottom largest faces of the sample, i.e. in the direction of the trapped flux (∥ $z$). The model does not take into account the penetration of the transverse field in the $x$-direction, i.e. the direction orthogonal to both the trapped flux and the transverse field. Surprisingly, the agreement between the modelled and the experimental data is very satisfactory. One possible explanation might be related to the *anisotropy* of flux penetration in the cross flux regime, as discussed below.

Using magneto-optical imaging, Indenbom *et al*. studied the geometry of penetration of the magnetic field $H \parallel c$-axis perpendicular to the largest surface of a thin YBCO single crystal in the presence of an applied longitudinal field $H \parallel ab$.[80] These results showed clearly that the flux lines ∥ $c$ penetrate and escape only *in the direction parallel the applied in-plane field*, in agreement with theoretical predictions of flux diffusion for this geometry.[81] Note that a similar behavior was also observed recently in (Bi,Pb)-2212 single crystals at low temperature.[82,83] The anisotropy in flux penetration can be explained qualitatively.[82,84] From a local point of view, the vortex lattice penetrating the sample moves easier along the pre-existing flux lines than across them in order to avoid cutting and reconnection processes.[85] From a global point of view, vortex motion across the existing flux lines induces currents flowing along them, which corresponds to a force-free configuration $\vec{B} \parallel \vec{J}$. The resulting "helical instability" of the flux line lattice[86] usually leads to much higher critical current densities than those determined by pinning ($\vec{B} \perp \vec{J}$),[80] i.e. better magnetic shielding across the trapped flux direction.

The experimental configuration studied here (cf. Fig. 1) differs completely from the geometries described above. The melt-textured samples (of aspect ratio 1 : 3) are first magnetized along their shortest dimension (the $z$-axis) and then subjected to a transverse field $H_y$ parallel to their largest surface. If the anisotropy of flux penetration holds true, however, the transverse flux lines are expected to penetrate and escape mainly along the $z$ direction and *not across* the trapped vortices. In this scenario, the flux lines move almost exclusively in the $y$-$z$ plane, whence the two-dimensional model of magnetic field components $H_y$ and $H_z$ is expected to describe appropriately the main features of the crossed field effects. In the present case, it is emphasized that the usual "infinite slab" geometry with both crossed fields parallel to the slab surface[14,26,30] would be inadequate since such a configuration corresponds to a transverse field moving *across* the direction of the existing trapped field.

The simple geometry used in the model is expected to give rise to some *quantitative* disagreement with the magnetic moment measurements, since the closure of the current loops is completely neglected. Another reason for a quantitative disagreement is that the anisotropy in current density cannot be included in the model. Our approach is expected to be reasonably valid for the case of melt-textured (RE)BCO materials characterized by a small anisotropy ratio ($J_c(ab) / J_c(c)$ ~ 1.5 to 2), but is probably unsuitable for the more anisotropic Bi cuprates. It should be noted that existing models that incorporate flux line cutting effects provide a *quantitative* prediction of the collapse in magnetization similar to the model described here. In ref. 30 for example, the remaining magnetic moment after one complete cycle of transverse field of amplitude $H_{max}$ ~ $H_p$ is approximately 0.11 times the initial value $M_0$, which should be compared to the modelled data in Fig. 3, showing that $B_z$ ~ 0.10 $B_0$ for similar experimental conditions.

The final part of this section addresses the physical mechanism responsible for the suppression of the magnetic moment under the application of a transverse field.

**C. Comparison between the model and the experiment**

Comparison of Figs. 2 and 3 reveals a very satisfactory qualitative agreement between the experimental $M_z(H_y)$ decays and the modeled data. The physical mechanisms involved in the suppression of trapped flux by cycling the transverse field can be understood by considering the current distributions in Figs. 4 and 5 and the magnetic flux distribution data in Fig. 8.

Firstly, the structure of the initial trapped flux distribution (|| $z$-axis) was shown to be defined by (i) the inhomogeneous current distribution resulting from the $J_c(B)$ dependence in the material and (ii) the finite thickness of the sample. As the transverse field ($H_y > 0$) increases, flux penetration occurs from the top and bottom and largest faces of the sample. As a result, two thin layers located in the vicinity of the two large faces carry currents required to shield the increasing transverse field d$H_y$/d$t$ > 0. As a result, $M_z$ decreases since these layers can no

longer contribute to the magnetization ∥ z-axis. Further increase of the transverse field leads to a rotation of the plane of symmetry of the current distribution, which contributes to a further decrease in $M_z$. In the case of a field-dependent $J_c$, the penetration of the transverse field towards the sample centre increases the total magnetic induction and reduces the current density. This effect also contributes to a reduction of the magnetization $M_z$. Once the applied transverse field reaches $H_p$ (i.e. the value corresponding to full-penetration along y in the absence of trapped flux ∥ z), the sample is only *partially* penetrated along the y-direction: the pinned vortices along z effectively retard the transverse field.

When the transverse field starts to decrease ($H_{max} \to 0$), the top and bottom layers of the sample begin to carry currents that oppose the negative variation in magnetic flux caused by the decreasing transverse field $dH_y/dt < 0$. The modification of the current distribution at both upper and lower parts of the sample does not influence the central zone responsible for the magnetization $M_z$, which remains almost unaffected. When the transverse field is removed, the current front lines exhibit an intricate "inverted Z" shape and some magnetization remains along the positive y-direction. Since the magnetic flux lines are necessarily closed, they find a return path in the free space around the sample and produce an additional z-component to the trapped flux $M_z$ at its top surface. This component is positive for $y > 0$ and negative for $y < 0$, resulting in a distorted flux profile, as observed experimentally (cf. Fig. 8).

When the transverse field is decreased from 0 to $-H_{max}$, the currents tend to oppose further the decreasing field. These currents extend towards the central zone of the sample, which results in further rotation of the current front lines. The final part of the cycle ($-H_{max} \to 0$) essentially affects the current distribution in the top and bottom regions of the sample, which are no longer contributing to the z-axis magnetization; i.e. $M_z$ remains constant in this regime. When the transverse field is removed, the y- component of the magnetization is directed towards $y < 0$. As a result, the flux profile at the top surface of the sample becomes distorted in the opposite sense to that during the first zero crossing.

According to the model proposed here, the electromagnetic response of a sample subjected to small amplitudes of the transverse field cycles ($H_{max} \sim 0.2\ H_p$) follows the scenario described above, i.e. a *monotonic* decrease of the magnetization $M_z$ during cycling $H_y$. Such a behavior is in perfect qualitative agreement with our experimental data (cf. Fig. 2) but differs from other experiments[15] or models[15,30] in which the $M_z(H_y)$ data are expected to follow some kind of "butterfly" loop. Further evidence of the validity of the model proposed here is given by comparing the electromagnetic response in the initial paramagnetic and diamagnetic states of the sample (Fig. 9). These results show a *symmetric* suppression of the magnetic moment by the transverse field, as observed experimentally by Fisher *et al.*[14] The symmetry in the response is also predicted by the two-velocity hydrodynamic model[12,14] and variational approaches[30] but not by the double critical state model[14,87].

Finally, the model proposed here was used to evaluate the behavior of a sample subjected to a large number of sweeps ($N \sim 100$). Again, the experimental (Fig. 6 and ref. 17) and modeled data (Fig. 7) behave in a very similar manner and predict that the trapped induction $B_z$ vs. $N$ decreases as a power law $B_z \propto N^{-\alpha}$. The important implication is that the magnetic induction does not stabilize at a plateau, even after more than 100 cycles of the transverse field. This suggests that the trapped flux in a type-II superconductor subjected to an orthogonal AC magnetic field should completely vanish after a sufficiently long time. In the case of positive field sweeps with $H_{max} \sim (1/6) H_p$, the measured $\alpha$ exponent is close to 0.026; the induction after 100 sweeps is reduced to 91 % of its initial value and continues to decrease. It is instructive to extrapolate this power law behavior to larger values of $N$, if the power-law behavior still holds true. After $N \sim 5.10^6$ cycles, which would correspond to the application of a 60 Hz AC field for one day, the magnetic induction should reach $\sim 68$ % of its initial value $B_0$, although a cycle time of more than 422 years would be needed to reach 50 % of $B_0$. The main conclusion to be drawn from this investigation is that the "AC regime" stabilization limit observed in the literature[12] might involve a very large number of field sweeps. Accordingly, great care should be taken when performing magnetic measurements in orthogonal AC and DC fields for extended periods of time, since the AC field may change significantly the DC magnetic flux in the time interval over which the experiment is performed.

# 6. CONCLUSIONS

The properties of type-II superconductors have been analyzed in the crossed field configuration, both experimentally and numerically. The geometry studied includes finite-size effects, i.e. the sample has a finite size along both (crossed) magnetic field directions. The magnetic behavior of the sample has been predicted using a two-dimensional finite-element model with currents flowing perpendicularly to the sample cross-section and closing at infinity. In this approach, the electric field, $E$, and the current density, $J$, are parallel to each other and are always orthogonal to the magnetic field $H$. As a result, neither flux free configurations or flux cutting are involved. A constitutive law $E \propto (J/J_c)^n$ is used in the model with only one critical current density $J_c$. A striking result of this study is that this simple approach is able to reproduce qualitatively the main features of the experimental data.

The present study focuses the magnetization collapse in a superconductor pre-magnetized along its shortest direction ($z$-axis) and then subjected to a series of transverse field cycles. The model shows that the suppression of magnetization results primarily from the modification of the current distribution in layers of the sample that are *perpendicular* to the direction of the pre-magnetization. The time-varying transverse field $H_y$ eventually leads to a *rotation* of the plane of symmetry of the current distribution in such a way the sample opposes the variation in magnetic flux imposed by $H_y$. The key point is the finite-size effect, i.e. the distribution of current along the thickness of the sample. Such a current distribution cannot be predicted by models that are based on either infinitely long (infinite slab or cylinders) or thin (thin strips or platelets) sample geometries. The agreement between the model and experiment provides evidence that the collapse of magnetization can be successfully predicted without the need of the detailed knowledge of the dissipation mechanism (flux free configuration or flux cutting effects) in the temperature and magnetic field range investigated. If a $J_c(B)$ dependence is considered, the decay of the *magnitude* of current density caused by an applied large transverse field adds to the modification to the current density distribution in reducing the sample pre-magnetization.

The results obtained in the present study allow significant conclusions to be drawn for the particular case of large bulk melt-textured YBCO pellets pre-magnetized along their shortest direction (|| $c$-axis). It is shown how the field-dependence of the critical current density $J_c(B)$ affects: (i) the structure of the trapped-field distribution and (ii) the decay rate of trapped flux.

The application of a transverse field has a profound impact on the current distribution and alters the symmetry of the trapped flux profile. Modeled and experimental data predict that (i) the application of one transverse field cycle of small amplitude yields a monotonic decrease of the trapped flux and (ii) a large number of sweeps in transverse field causes the magnetic induction to decrease continuously following a power law whose exponent is related to the sweep amplitude. The model proposed in this study offers a simple way of predicting the electromagnetic behavior of YBCO pellets subjected to crossed magnetic fields.

# ACKNOWLEDGMENTS

Ph.V. is grateful to the FNRS for a travel grant. We thank Profs. R. Cloots, M. Dirickx, M. McCulloch and B. Vanderheyden for fruitful discussions and comments. We are grateful to D.M. Astill, S. Dorbolo, J. F. Fagnard, Ph. Laurent, M. Majoros, B. Mattivi, I. Molenberg, and Y. Shi, for their assistance in carrying out measurements, and to J.-P. Mathieu for preparing some of the samples. Part of this work was supported by the European Commission through the RTN 'SUPERMACHINES' contract (HPRN-CT-2000-036) and by the Région Wallonne (RW) through the 'VESUVE' contract. We also thank the FNRS, the ULg and the Royal Military Academy (RMA) of Belgium for cryofluid and equipment grants. Part of this work also supported by the Visitor's Program of the National High Magnetic Field Laboratory, which is supported by the U.S. National Science Foundation.

**TABLES**

TABLE I: dimensions and full-penetration fields $H_p$ of the melt-textured single-domain samples

| Sample | Shape | Thickness ∥ c-axis | Surface ∥ ab planes | $\mu_0 H_p$ ∥ c | $\mu_0 H_p$ ∥ ab |
|---|---|---|---|---|---|
| HB1 ($YBa_2Cu_3O_7$) | Cylindrical puck | 9.5 mm | disk of 23 mm diameter | 0.32 T (77 K) | 0.25 T (77 K) |
| HB2 ($DyBa_2Cu_3O_7$) | Parallelepiped | 0.56 mm | square 1.78 x 1.78 mm | 0.10 T (80 K) | 0.07 T (80 K) |

TABLE II: parameters used for modeling

| | Fixed parameters | Variable parameters |
|---|---|---|
| Geometry | Aspect ratio $z_0 / y_0 = 1/3$<br>Magnetizing field $H_z$ ∥ z-axis<br>Transverse field $H_y$ ∥ y-axis | ------- |
| Material properties | $n$ = constant (= 21) | $J_c = J_{c1}$ or<br>$J_c = J_{c1} (1 + 4|B|/B_p)^{-1}$<br>where $B_p = \mu_0 J_{c1} (y_0 / 2)$ |
| Applied fields | Ramp $\mu_0 H_z$ up to $2B_p$ for 0.1s<br>Ramp $\mu_0 H_z$ down to 0 for 0.1s<br>Wait for 10 s, then apply the transverse fields | Several amplitudes of transverse fields $H_y$<br>Several sweep rates $dH_y/dt$ |

# FIGURE CAPTIONS

FIG. 1. (a) Schematic illustration of the experimental configuration used for "crossed field" measurements. (b) Geometry used for the two-dimensional model: the sample is of cross-section $y_0 \times z_0$ and is infinite in the $x$ direction. In all cases, the sample is pre-magnetized along the $z$-axis and the transverse field is applied parallel to the $y$-axis.

FIG. 2. Measured central magnetic induction $\| z$ (sample HB1) during the application of one cycle of transverse field $\| y$, as shown in the inset. The induction is normalized with respect to its initial value $B_0$. The transverse field is normalized with respect to the full penetration field $H_p$, which is determined experimentally. The amplitudes of the cycles $H_{max} / H_p$ are 0.15, 0.30, 0.45, 0.98, 1.18 and 1.53.

FIG. 3. Results of the numerical model of the central magnetic induction $\| z$ during the application of one cycle of transverse field $\| y$, as shown in the inset. The induction is normalized with respect to its initial value $B_0$. The transverse field is normalized with respect to the full penetration field $H_p$. The amplitudes of the cycles $H_{max} / H_p$ are 0.15, 0.30, 0.45, 0.98, 1.18 and 1.53.

FIG. 4. Modeled data of the current density distribution $J_x(y, z)$ within the cross-section of the sample during one cycle of the transverse magnetic field of amplitude 0.5 $H_p$. A field-dependent $J_c(B)$ is assumed. Bottom: scale of current density $J_x$ (expressed in $10^3$ A/cm²). The arrow indicates the direction of a positive transverse field.
Right: schematic diagram showing the times at which the current density distributions are determined.

FIG. 5. Modeled data of the current density distribution $J_x(y, z)$ within the cross-section of the sample during one half-cycle of the transverse magnetic field:
(a) constant $J_c$ and $H_{max} = 0.5\ H_p$ (b) constant $J_c$ and $H_{max} = 1.5\ H_p$ (c) field-dependent $J_c(B)$ and $H_{max} = 1.5\ H_p$. The times (0) - (4) in the cycles correspond to those defined in Figure 4.
(d) Modeled data of the average magnetization $M_z$ during one cycle of the transverse field, assuming either a constant (●) or a field-dependent (○) critical current density $J_c$.

FIG. 6. Log-log plot of the measured central magnetic induction (sample HB1) at the end of each positive transverse field cycle for three different transverse field sweep amplitudes $H_{max}$. The plain lines are power law fits $B_z \sim N^{-\alpha}$, with values of $\alpha$ given in the inset.

FIG. 7. Log-log plot of the modeled central magnetic induction at the end of each positive transverse field cycle for three different transverse field sweep amplitudes $H_{max}$. The plain lines are power law fits $B_z \sim N^{-\alpha}$, with values of $\alpha$ given in the inset.

FIG. 8. (a) MOI image of the initial trapped flux above the top surface of sample HB2. The arrow indicates the transverse field direction. The intersection of the two orthogonal dashed lines defines the sample center ($x = 0$, $y = 0$).

(b) Comparison of the flux profiles $|B_z(y)|$ measured above the top surface of sample HB2 along the $y$ direction before and after the application of the transverse field, as shown in the inset. The field amplitude $H_{max} = 0.48\ H_p$.

(c) Modeled data of the distribution of $|B_z(y)|$ along the $y$ direction at the top surface before and after the application of the transverse field, as shown in the inset. The field amplitude $H_{max} = 0.50\ H_p$. The magnetic induction is normalized with respect to the maximum trapped field value.

FIG. 9. Modeled data of the $M_z(H_y)$ curves obtained for the "paramagnetic" and the "diamagnetic" initial states resulting from the magnetization sequences $M_z(H_z)$, as shown schematically in the inset. In all cases, the magnetizing field $H_z$ is applied continuously during the application of the tranverse field $H_y$.

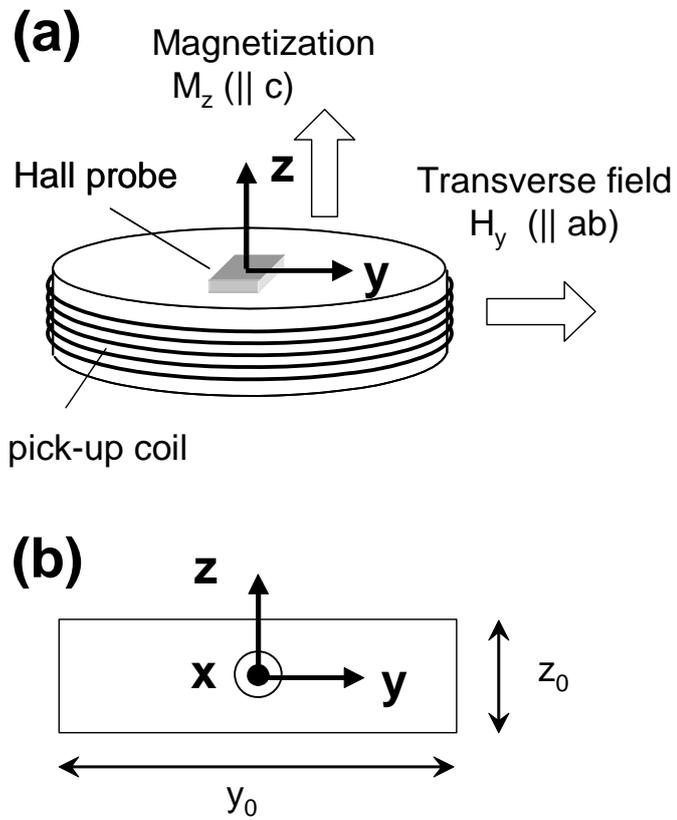

**Figure 1**

**Vanderbemden *et al.***

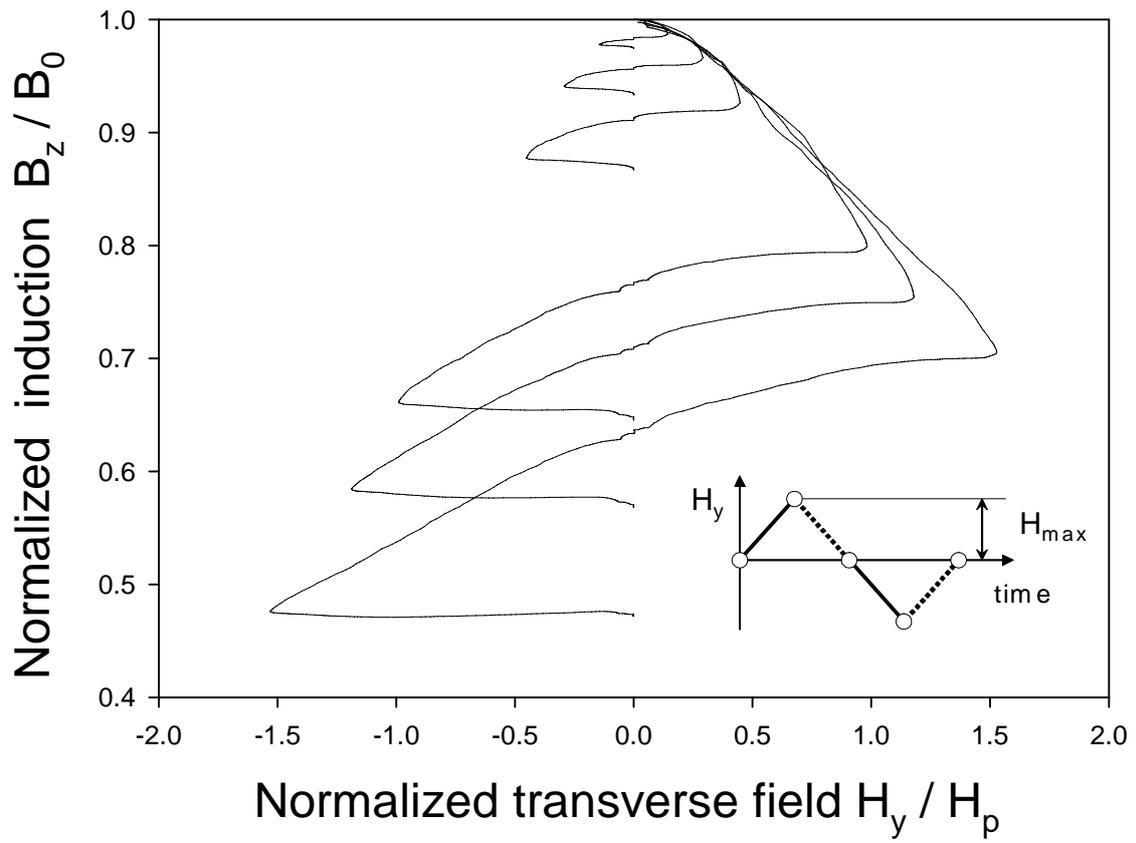

**Figure 2**

**Vanderbemden *et al.***

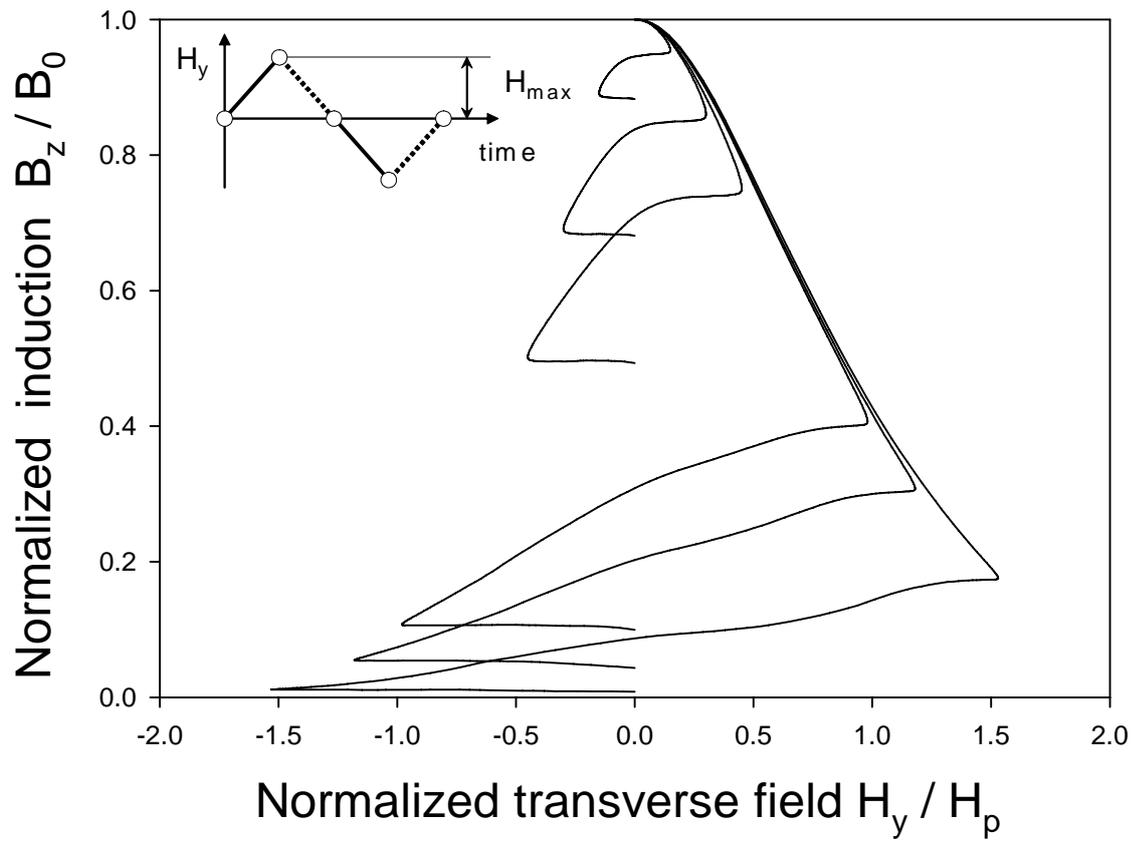

**Figure 3**

**Vanderbemden *et al.***

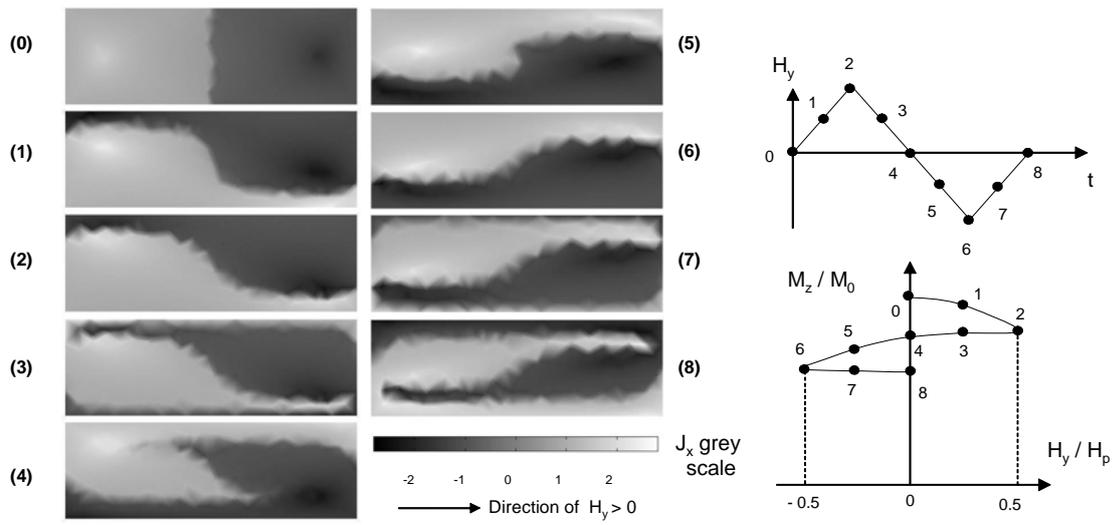

**Figure 4**

**Vanderbemden *et al.***

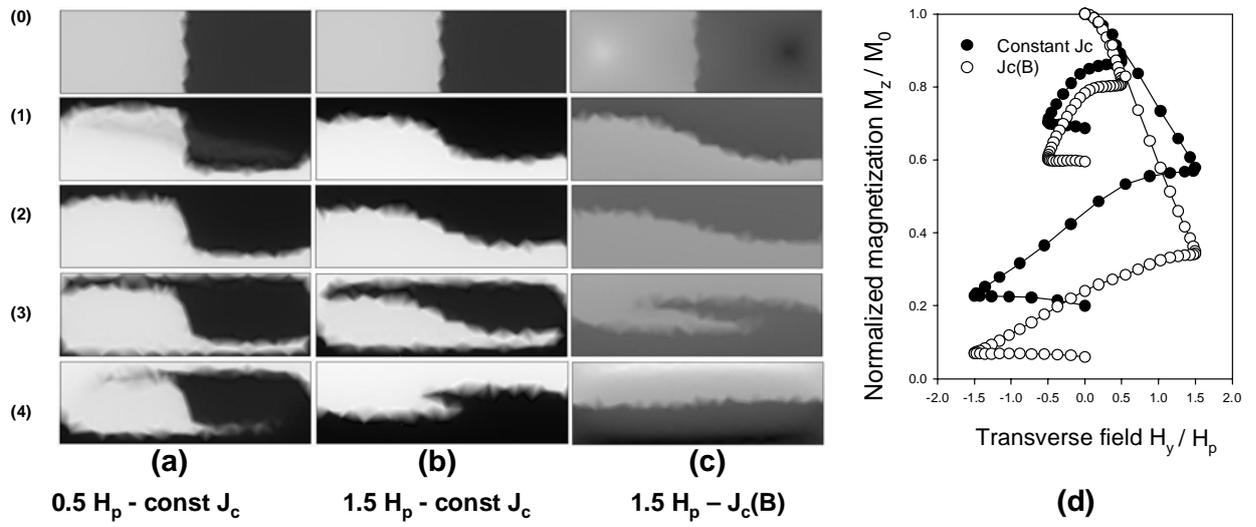

**Figure 5**

**Vanderbemden *et al.***

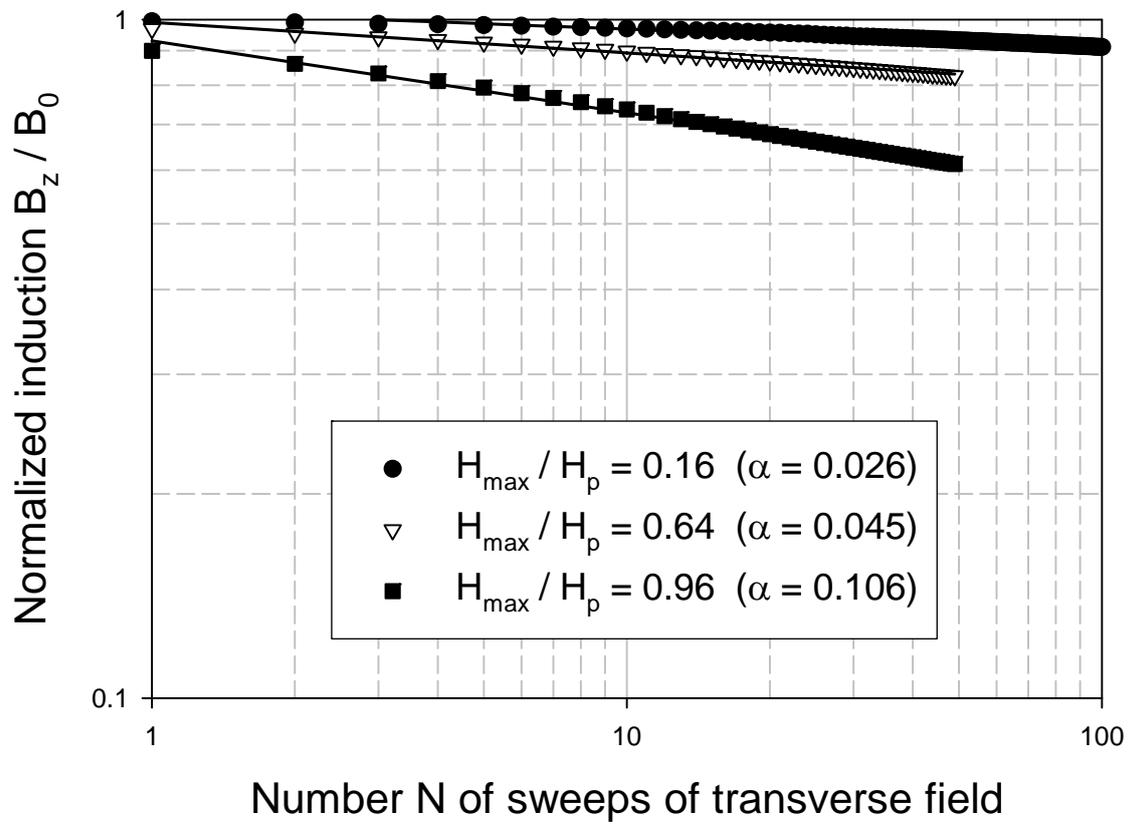

**Figure 6**

**Vanderbemden *et al.***

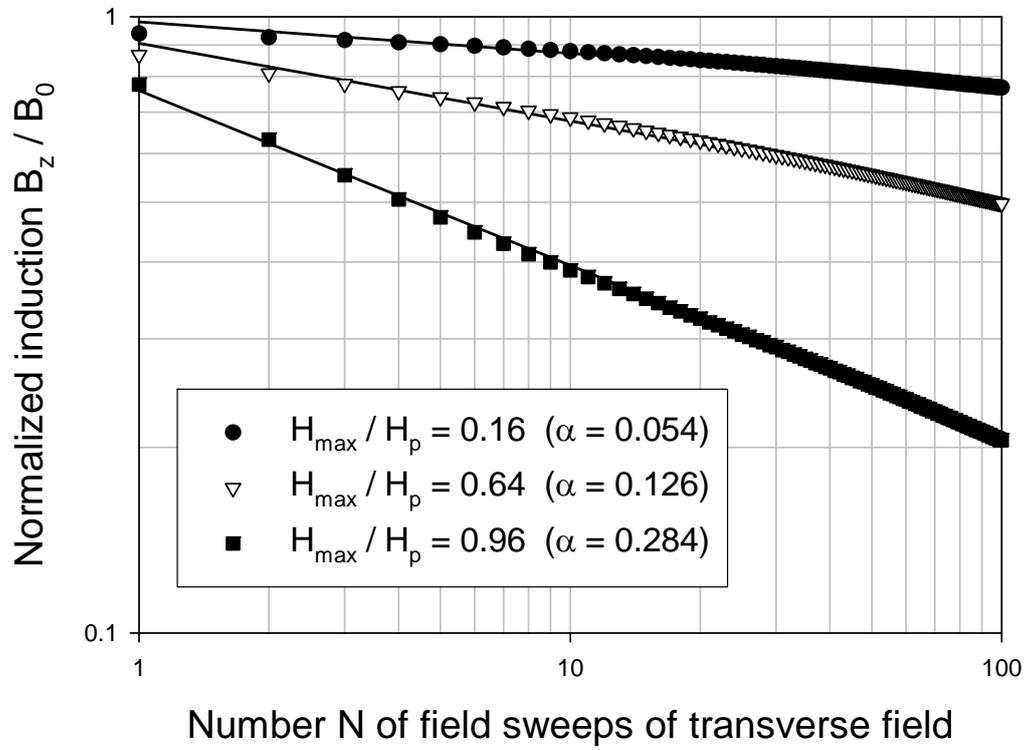

**Figure 7**

**Vanderbemden *et al.***

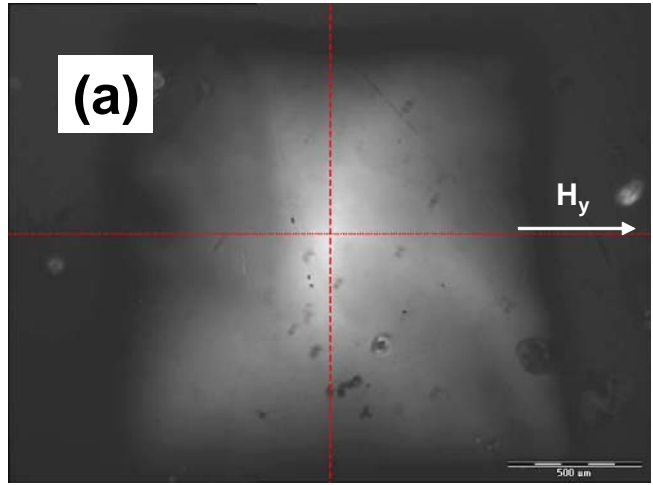

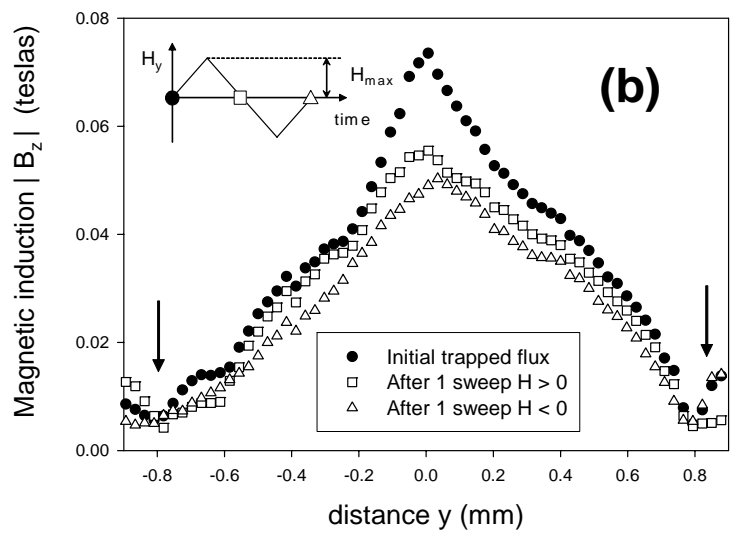

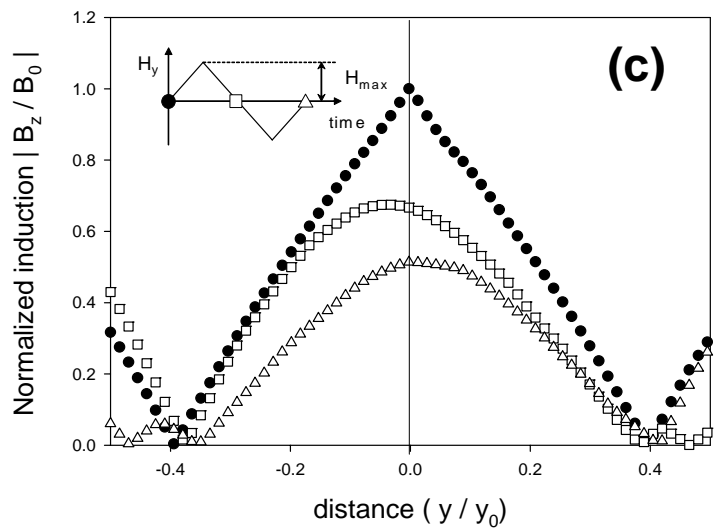

**Figure 8
Vanderbemden *et al.***

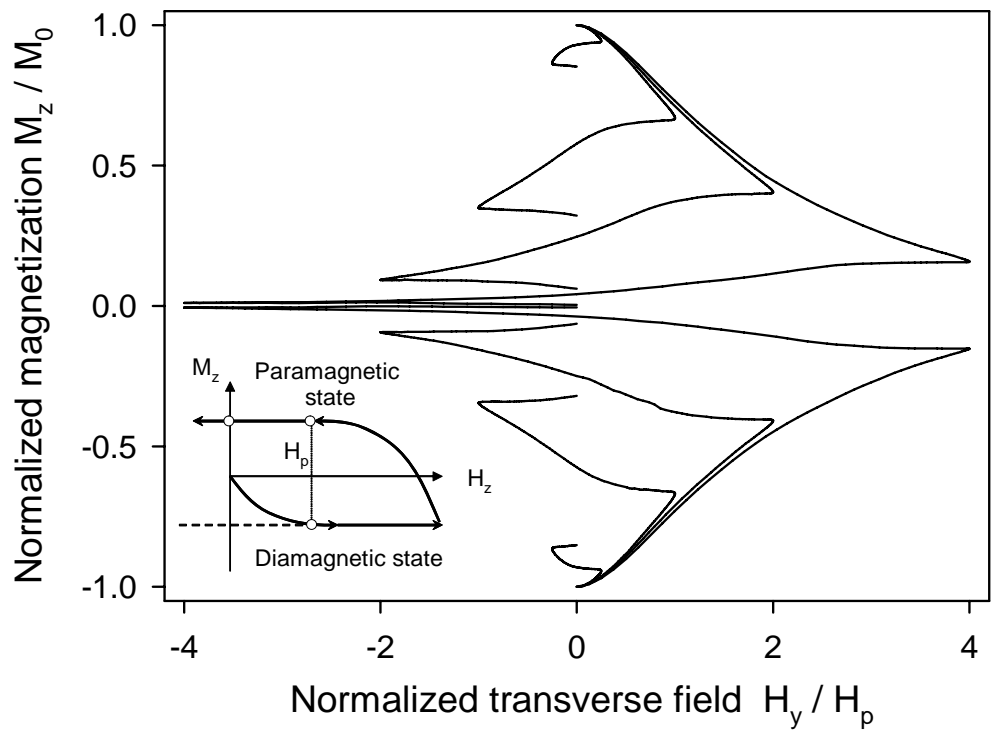

**Figure 9**

**Vanderbemden *et al.***